\documentclass[aps,prb,twocolumn]{revtex4-1}
\usepackage[T1]{fontenc}
\usepackage{graphicx}
\usepackage{calc}
 \usepackage{amsmath}

\newcommand{\Sj}{{\bf S}_j}
\newcommand{\Sk}{{\bf S}_k}
\newcommand{\Sl}{{\bf S}_l}
\newcommand{\Sm}{{\bf S}_m}

\newcommand{\brjk}{\langle j,k \rangle}

\newcommand{\Wq}{{\bf W}_{\bf q}}
\newcommand{\Wdagq}{{\bf W}^\dagger_{\bf q}}
\newcommand{\Wnegq}{{\bf W}_{-\bf q}}
\newcommand{\Wdagnegq}{{\bf W}^\dagger_{-\bf q}}

\newcommand{\omegaq}{ \omega ({\bf q})}
\newcommand{\ruj}{{\bf r}_j}
\newcommand{\ruk}{{\bf r}_k}
\newcommand{\rul}{{\bf r}_l}

\begin{document}

\title{Independent Bond Fluctuation Approximation to the Ground State of Quantum Antiferromagnets}
\author{Christian Rischel } \affiliation{Novo Nordisk, Novo Nordisk Park, 2760
M\aa l\o v, Denmark} \email{cris@novonordisk.com}

\begin{abstract}
A simple approach to estimation of the ground state energy of quantum antiferromagnets is developed, based on the approximation that quantum fluctuations around different bonds are independent. The ground state energy estimates are as good as spin wave theory or slightly better. A canonical transformation of the spin operators to generate bond quantum fluctuations is devised and applied to the classical ground state of the $S=\frac{1}{2}$ Heisenberg model on the square lattice. This simple picture of quantum spin fluctuations might be useful in more complex models. The resulting nearest neighbor and next-nearest neighbor correlations can be used in an alternative derivation of spin waves in the Heisenberg model, giving an accurate dispersion as well as raising and lowering operators.
\end{abstract}

\maketitle

The ground state energy of antiferromagnets are lowered relative
to the classical energy of the N\'eel state by quantum effects.
For the Heisenberg model, spin-wave theory\cite{SpinWave} gives a
relatively good quantitative prediction of the shift. Here, I show that a simple
estimation of the quantum contribution at individual bonds gives
surprisingly good predictions of the ground state energy of XXZ
and XY models. I also devise an explicit transformation of the
spin operators that generates fluctuations around bonds when
applied to the classical ground state. The energy of the resulting
state is calculated on the square lattice and found to be close to the simpler estimate.

\section{The independent bond fluctuation picture}

The quantum contribution to the ground state energy can mathematically be understood as arising
from the competition between diagonal and off-diagonal terms in
the Hamiltonian. The lowest energy of the diagonal part is found
when the system is localized in a particular basis state, while
minimizing the energy from the off-diagonal part requires a
superposition of basis states. The Hamiltonian for the nearest neighbour XXZ model
is
\begin{eqnarray}
\nonumber H & = & J \sum_{\brjk} [\alpha(s_j^x s_k^x+s_j^y
s_k^y)+s_j^z s_k^z ] \\ & = & J \sum_{\brjk}
[\frac{\alpha}{2}(s_j^+s_k^- + s_j^-s_k^+) + s_j^z s_k^z] ,
\label{eq:XXZHamil}
\end{eqnarray}
where the sum runs over nearest neighbors. Here, I will only
discuss values of the anisotropy parameter $\alpha$ in the range
$0 < \alpha \leq 1$. The Heisenberg model is obtained with
$\alpha=1$. If we take as basis the states where each spin $i$ is
in an eigenstate of $s_i^z$, the '$zz$'-terms in
(\ref{eq:XXZHamil}) are diagonal and the '$+-$'-terms are
off-diagonal.
I now restrict the discussion to bipartite lattices, which
can be divided into two sublattices with no interactions between
spins on the same sublattice. The minimum energy of the diagonal
terms is for antiferromagnets ($J>0$) found with all spins on one
sublattice having $s_j^z =+S$ ('up'), and all spins on the other pointing
the opposite way ($s_k^z=-S$) ('down'), where $S$ is the length of one
spin. This is also the classical ground state. The energy is
$-JS^2$ per bond. The off-diagonal terms all have magnitude
$\alpha JS$. They couple the classical ground state to excited basis states in which one 'up' spin has been lowered to $s_j^z = S-1$ and a neighboring 'down' spin has been incremented to $s_k^z = -S+1$. If the number of nearest neighbors in denoted by $Z$, each of these two spins have diagonal couplings to $Z-1$ other spins of length $S$. In the excited basis states, this diagonal energy has increased by $2JS(Z-1)$. Even for $\alpha=1$, this increase in diagonal energy is substantially higher than the
off-diagonal coupling, so we may expect the zero-point
fluctuations to be relatively small. I therefore make the
approximation that \emph{quantum ground state fluctuations about different
bonds are uncorrelated}. To calculate the depression of ground
state energy per bond we only need to take into account the basis
state that the Hamiltonian of one bond couples to, with the
diagonal and off-diagonal terms given above.
\begin{figure}
\includegraphics[width=\linewidth]{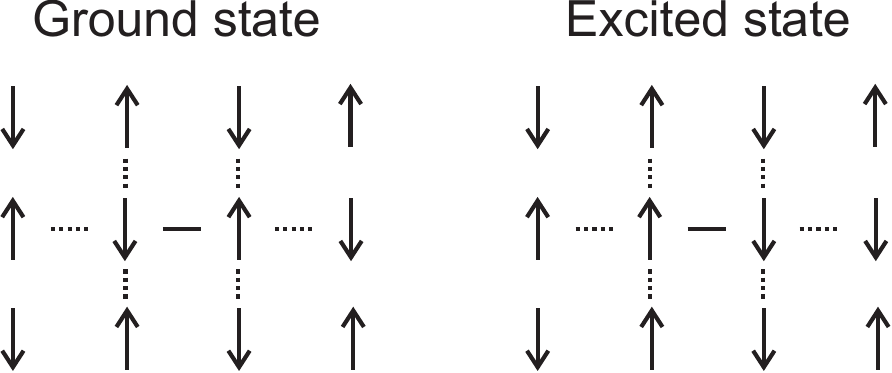}
\label{fig:SquareTwoLevelSystem}
\caption{The two-level system exemplified on the square lattice. Spin exchange is considered along the bond shown with solid line, while only the diagonal ($s_j^z s_k^z$) energy terms are considered along the bonds shown with dotted lines.}
\end{figure}
This is a simple two-state system with the states illustrated in Figure \ref{fig:SquareTwoLevelSystem} for$S=\frac{1}{2}$ on the square lattice.
The ground state energy per bond $E_\mathrm{XXZ}$ can
then be found as the lowest eigenenergy of the matrix
\begin{equation}
\left(
\begin{array}{cc}
-J S^2 & \alpha JS \\
\alpha JS & -JS^2 + 2JS(Z-1)
\end{array}
\label{eq:2x2Matrix}
\right)
\end{equation}
which is
\begin{equation}
 E_\mathrm{XXZ} = -JS^2+JS(Z-1) - JS\sqrt{(Z-1)^2+\alpha^2}
 \label{eq:HeisE0}
\end{equation}
If we write the corresponding eigenvector in the form
\begin{equation}
\left(
\begin{array}{c}
1 \\
- \eta
\end{array}
\right)
\label{eq:TwoStateVector}
\end{equation}
the relative amplitude of the basis state with exchanged spins is then
\begin{equation}
\eta = \frac{1}{\alpha} (\sqrt{(Z-1)^2+\alpha^2}-(Z-1) \label{eq:etaexact})
\end{equation}
The nearest neighbour XY model has the Hamiltonian
\begin{equation}
H_\mathrm{XY} = J \sum_{\brjk} [s_j^x s_k^x+s_j^y s_k^y ]
\end{equation}
It can easily be handled by the
method developed here if it is considered the 'XZ' model in the
chosen basis\cite{XYSpinWave}:
\begin{eqnarray}
H_\mathrm{XZ} & = & J \sum_{\brjk} [s_j^x s_k^x+s_j^z
s_k^z ] \\
\nonumber & = & J \sum_{\brjk} [\frac{1}{4}(s_j^+s_k^- +
s_j^-s_k^+ +s_j^- s_k^- + s_j^+s_k^+) + s_j^z s_k^z]
\end{eqnarray}
The ground state and the diagonal energies are the same as for the XXZ model with $\alpha=0.5$, but in
addition to the '$+-$'- and $'-+$'-terms found in
(\ref{eq:XXZHamil}), '$++$'- and $'--'$-terms now appear. However,
these new terms destroy the state with lowest diagonal energy so
they can be disregarded for the calculation of the ground state
energy by the present method. The remaining off-diagonal terms are $JS/2$, so the
estimated ground state energy per bond becomes
\begin{equation}
E_\mathrm{XY} = -JS^2+JS(Z-1) - JS\sqrt{(Z-1)^2+1/4}
\label{eq:XYE0}
\end{equation}
It is simple to do the calculation for other anisotropic models with antiferromagnetic spin order on bipartite lattices.
Obviously, the strongest interactions should be along the
$z$-direction. Models with interactions between e.g. $S=\frac{1}{2}$ and $S=1$ spins can also be handled by the same general method, although new expressions for both diagonal and off-diagonal terms must be devised.

\begin{table}[t]
\begin{minipage}{\columnwidth}
\centering
\begin{tabular}{|l|c|c|c|c|}
\hline Model & Reference & Spin-wave & This work & 1st order\\
\hline \hline 1D Heisenberg & -0.4431\footnote{Exact\cite{Hulthen}.} & -0.4315 & -0.4571 & -0.5\\
\hline Honeycomb Heis. & -0.3629\footnote{Series expansion\cite{Oitmaa1992}.} & -0.3549 & -0.3680 & -0.375\\
\hline Square Heisenb. & -0.3347\footnote{QMC\cite{Sandvik1997}.} & -0.329 & -0.3311 & -0.3333\\
\hline Cubic Heisenb. & -0.2998\footnote{Self-consistent mean-field\cite{SelfConsistent}.} & -0.299 & -0.2995 & -0.3 \\
\hline 1D XY & -0.318\footnote{Exact\cite{XYExact}.}  & -0.299 & -0.3090 & -0.3125\\
\hline Square XY & -0.2745\footnote{Correlated basis functions\cite{XYAbInitio}.} & -0.27 & -0.2707 & -0.2708\\
\hline Cubic XY & -0.2640$^{ f}$ & -0.26 & -0.2625 & -0.2625\\
\hline
\end{tabular}
\end{minipage}
\caption{\label{tab:E0} Ground state energy ($J=1$) per bond of
$S=\frac{1}{2}$ antiferromagnetic models calculated by different
methods. In all cases the classical energy is -0.25.}
\end{table}

Table \ref{tab:E0} lists the ground state energies for the
$S=\frac{1}{2}$ Heisenberg and XY models on selected lattices,
with $J=1$. Shown are estimates of the values from exact methods
or accurate numerical calculations (see footnotes for references),
the results from first order spin wave
theory\cite{SpinWave,XYSpinWave} and the values found by equation
(\ref{eq:HeisE0}) or (\ref{eq:XYE0}). These approximations are
seen to be as good as the spin wave results, in most cases even
slightly better (calculations on the square lattice XXZ model, $\alpha<1$, follow
below). The good agreement suggests that the simple picture of
uncorrelated bond fluctuations is a reasonable representation of local
correlations in the true ground state. The table also lists the values of the first-order expansion of (\ref{eq:HeisE0}) in $\alpha^2/(Z-1)^2$:
\begin{equation}
 E_\mathrm{XXZ} \approx -JS^2 - JS\frac{\alpha^2}{2(Z-1)} 
 \label{eq:HeisE1}
\end{equation}
and similarly for (\ref{eq:XYE0}). This result could also have been obtained by second-order perturbation theory. The numbers are also seen to be quite accurate, with the exception of the 1D Heisenberg model.

The energy estimates (\ref{eq:HeisE0}) and (\ref{eq:XYE0}) do not take into account
correlations between different bonds. In fact, the diagonal energy change $2JS(Z-1)$ used in the derivation of (\ref{eq:2x2Matrix}) no longer holds when spins in a neighboring bond are exchanged.
For this reason, one would expect the approximation to be best when the amplitude of spin exchange is small. Expansion of (\ref{eq:etaexact}) to first order gives $\eta \approx \alpha^2/2(Z-1)$, and in fact the results in Table \ref{tab:E0} are better the higher the number of nearest neighbours. For cases with $S > \frac{1}{2}$, exchange of neighboring spins does not flip the neighboring spins completely. Hence, the relative change of the diagonal energy is smaller, and the approximation can be expected to work better for larger values of $S$. Indeed, for the 1D $S=1$ Heisenberg model, equation (\ref{eq:HeisE0}) with $J=1$ gives the ground state energy $-\sqrt{2} \sim -1.414$, close to the very accurate numerical estimate -1.401 \cite{WhiteHuse}. Results in higher dimensions and/or for higher values of $S$ are expected to be even better.

\section{Staggered magnetization}

The staggered magnetization (or sublattice magnetization) can be calculated in the independent bond fluctuation picture by considering that a given spin either can be in its' ground state configuration or in one of the $Z$ states where spin has been exchanged with a neighbouring site. In order to calculate the probability of the different states, the values derived from (\ref{eq:TwoStateVector}) must be normalized by a factor $1+Z\eta^2$ in order for the probabilities to sum to 1. The staggered magnetization then becomes
\begin{equation}
M = \frac{S + Z\eta^2 (S-1)}{1 + Z\eta^2}
\label{eq:StagM}
\end{equation}
Table \ref{tab:StagM} lists the results of this formula for the  $S=\frac{1}{2}$ XXZ model on the square lattice at different values of $\alpha$, together with more accurate reference values. It is clear that the independent bond fluctuation picture does a quite poor job of calculating the staggered magnetization, in particular at values of $\alpha$ close to 1. This shows that only {\em local} correlations in the ground state are captured well. It can be noted that other approaches based on expansions of local interactions suffer from similar problems\cite{XXZGroundState}.
\begin{table}
\begin{minipage}{\columnwidth}
\centering
\begin{tabular}{|l|c|c|c|c|}
\hline $\alpha$ & 0.5 & 0.8 & 0.9 & 1.0 \\
\hline \hline Reference value & 0.471  & 0.417 & 0.386 & 0.307 \\
\hline \hline Equation (\ref{eq:StagM}) & 0.473 & 0.436 & 0.420 &  0.405 \\
\hline
\end{tabular}
\end{minipage}
\caption{Calculated sublattice magneization for the S=$\frac{1}{2}$ XXZ model on the square lattice. Reference values are from series expansion
\cite{Zheng1991}. \label{tab:StagM}}
\end{table}

\section{Surface energy at magnetic zone boundaries}

In the independent bond fluctuation picture, it is simple to estimate quantum effects on the surface energy between magnetic zone boundaries. A particularly simple case is the lowest energy of a localized, flipped spin on an antiferromagnetic background. In the ground state, all bonds have the energy (\ref{eq:HeisE0}). After flipping a single spin, the $Z$ bonds connected to that spin will have energy $JS^2$. However, the other bonds connected to the nearest neighbors are also affected, since the diagonal energy opposing fluctuation of these bonds decreases to $2JS(Z-2)$. In the lowest energy state, the fluctuation becomes stronger and the energy is lowered to
\begin{equation}
-JS^2 + JS(Z-2) - JS \sqrt{(Z-2)^2+\alpha^2}
\label{eq:SpinflipAFnnE}
\end{equation}
\begin{figure}
\includegraphics[width=5cm]{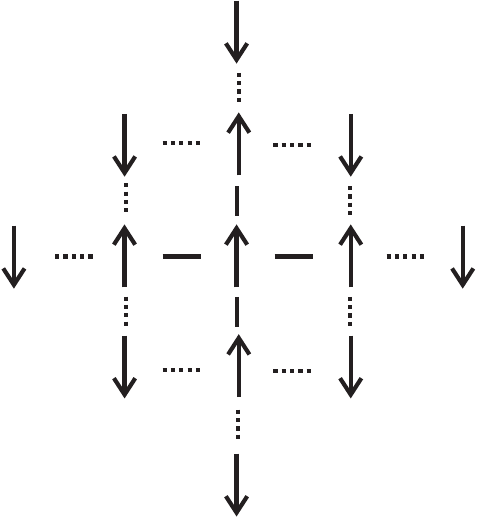}
\label{fig:SquareOneFlip}
\caption{Flip of a single spin on the antiferromagnetic square lattice. Relative to the ground state, the central spin has been flipped. The bonds shown with solid lines have an increase in energy, while the bonds shown with dotted lines have a (small) decrease in energy.}
\end{figure}
Taking the $S=\frac{1}{2}$ Heisenberg model on the square lattice as example, this energy is $-0.3680 J$, so the energy of each bond is lowered by a relatively modest $0.0369 J$. However, as illustrated in Figure \ref{fig:SquareOneFlip} there are 12 such bonds, so the total contribution is $-0.4427 J$. The total energy of a flipped spin in this case becomes
\begin{equation}
(4 \times 0.25 - 0.4427 - 4\times -0.3311) J = 1.8818 J
\end{equation}
It is a bit counterintuitive that the energy cost of flipping a spin is lower than the classical value $2J$, since we are breaking antiferromagnetic bonds with a stronger interaction than the classical value, but the cumulative effect on neighboring bonds is found to be the larger quantum effect. The analysis is easily carried to cases with two spins flipped at next-nearest neighbor positions. The results for the $S=\frac{1}{2}$ square lattice Heisenberg model show a very weak attraction in the axial position, probably much lower than the precision of the model, and a repulsion on the order of $0.3J$ in the diagonal position.

 This approach more generally predicts locally enhanced antiferromagnetic interactions at interfaces between ferromagnetic and antiferromagnetic zones.  This could have interest in understanding {\em e.g.} the Hubbard model under certain conditions, where  antiferromagnetic and ferromagnetic tendencies compete.

\section{Transformation of spin operators}

The derivation of (\ref{eq:HeisE0}) consists of taking the
classical ground state and introducing uncorrelated perturbations
at each bond:
\begin{equation}
\prod_{\brjk} [1-\eta s^-_j s^+_k /2S] | 0
\rangle \label{eq:IBF}
\end{equation}
where $|0\rangle$ is the classical ground state and the product
runs over all pairs of nearest neighbours with index $j$ counting
'up'-spins and $k$ counting 'down'-spins. Clearly, $\eta$ is the
amplitude of the state with spins exchanged at one bond, given by (\ref{eq:etaexact}). This is very similar to the Local Ansatz approximation developed by Stollhoff and Fulde \cite{Stollhoff1977}, and also close to the
Coupled Cluster Method, although the ground state energy estimates obtained here are better than those obtained by analytical CCM \cite{FarnellBishop2006}.

The relation $[s^+_j ,s^-_j]=2s^z_j$ holds generally for spin
operators. In the following, the discussion is limited to the case $S=\frac{1}{2}$,
where $[s^+_j ,s^-_j]|0\rangle = |0\rangle$ if spin $j$ belongs
to the sublattice pointing up. Therefore, one may treat the spin
operators as bosonic with $s_j^-$ as creation operator and $s_j^+$
as destruction operator. This is one of the ways to express the
basic idea behind spin wave theory\cite{HolsteinPrimakoff}. On the
'down'-sublattice, $s_k^-$ destroys the classical ground state and
$[s^-_k ,s^+_k]|0\rangle = |0\rangle $. It might be interesting to
find operators which destroy the state (\ref{eq:IBF}) instead, since it has
an improved description of local quantum correlations in the XXZ
model. $s^+_j$ does the job to zeroth order in $\eta$ on the
'up'-sublattice, but it can be seen from (\ref{eq:IBF}) that terms
of amplitude $-\eta$, with a spin pointing up on one of the
neighboring sites, still remain. These terms can be neutralized
by adding operators that construct the same terms with amplitude
$+\eta$:
\begin{equation}
s^{+}_j{}' = s^+_j + \eta \sum_{k \in {\cal N}_j} s^+_k
\end{equation}
where ${\cal N}_j$ is the set of nearest neighbours to spin $j$.
This construction is not ideal, since $s^+_j{}'$ and $s^-_k{}'$ on
neighboring sites $(j,k)$ don't commute. A more satisfactory
operator is
\begin{equation}
t^{+(1)}_j = s^+_j \pm 2\eta s^z_j \sum_{k \in {\cal N}_j} s^+_k
\label{eq:t1def}
\end{equation}
which still destroys the state (\ref{eq:IBF}) to first order in $\eta$, but
also commutes with $t^{-(1)}_k$ on neighboring sites to first
order in $\eta$. The positive sign in (\ref{eq:t1def}) holds on
'up'-sites, the negative sign on 'down'-sites. I now introduce the
operator
\begin{equation}
L = \eta \sum_{\brjk} \left( s^-_j s^+_k - s^+_j s^-_k  \right)
\label{eq:LDef}
\end{equation}
where the sum runs over all pairs of nearest neighbors on the
lattice, and spins of index $j$ are on the 'up'-sublattice in
$|0\rangle$ and spins of index $k$ are neighbors to $j$ on the 'down'-sublattice.
$L$ is obviously anti-Hermitian ($L^\dagger = -L$), so the
operator transformation
\begin{eqnarray}
t^{\pm}_j & = & e^{-L} s^{\pm}_j e^{L} \label{eq:LTranspm} \\
t^z_j & = & e^{-L} s^z_j e^{L} \label{eq:LTransz}
\end{eqnarray}
conserves all commutation relations; the $t$ operators obtained by
this transformation are \textit{bona fide} spin-$\frac{1}{2}$
operators, albeit delocalized on the lattice. Equation
(\ref{eq:LTranspm}) can be expanded
\begin{eqnarray}
t^+_j & = & s^+_j + [L,s^+_j]+\frac{1}{2!}[L,[L,S^+_j]] + ... \label{eq:TExpand} \\
& = & s^+_j + 2\eta s^z_j \sum_{k \in {\cal N}_j} s^+_k + {\cal
O}(\eta^2)
\end{eqnarray}
showing that $t^{+(1)}_i$ is exactly $t^+_j$ expanded to first
order. $L$ can alternatively be written
\begin{eqnarray}
L & = & 2i\eta \sum_{\brjk} \left( s^x_j s^y_k - s^y_j s^x_ {}\right) = 
2i \eta  \sum_{\brjk} \left( {\bf s}_j \times {\bf s}_k \right)^z \nonumber \\
& = & -\frac{\eta}{\alpha} [ H, \sum_j s^z_j - \sum_k s^z_k ]
\label{eq:LAlternative}
\end{eqnarray}
The asymmetric form of $L$ distinguishes the transformation developed here from the Local Ansatz method\cite{Stollhoff1977}, and is specific for antiferromagnetic interactions.

The $t^+_j$ operator destroys the state (\ref{eq:IBF}) to
first order, and it {\em exactly} destroys the state
$e^{-L}|0\rangle$. This state is simply the classical ground state
in the (slightly) delocalized spin operators defined by
(\ref{eq:LTranspm}-\ref{eq:LTransz}). To first order in $\eta$, it displays the same bond fluctuations as (\ref{eq:IBF}), but it is correctly normalized and allows for higher order calculations.

\section{Numerical ground state energies}
Estimates of ground state energies can be calculated by expanding
$ E^0(L) = \langle 0 | e^{L}H_{jk}e^{-L}|0 \rangle$
to a given order with the same expansion as in (\ref{eq:TExpand}) and calculating the value of terms with only $s_z$ operators  and
calculating the value of terms with only $s_z$ operators  (since $\langle 0| s^x_j|0\rangle = \langle 0|s^y_j|0\rangle =0 $). Here, $(j,k)$ is any pair of nearest neighbors. I have obtained
an approximate numerical value of $E^0(L)$ for XXZ models on the square
lattice, by expanding $e^{L}H_{jk}e^{-L}$ to high orders in a
symbolic computation. In $L$, terms were included from a part of
the lattice around the $(j,k)$-pair large enough to avoid any
finite-size effects.

\begin{table}[b]
\begin{minipage}{\columnwidth}
\centering
\begin{tabular}{|l|c|c|c|c|}
\hline $\alpha$ & 0.5 & 0.8 & 0.9 & 1.0 \\
\hline \hline Reference energy  & -0.2708 & -0.3037 & -0.3183 & -0.3362 \\
\hline \hline $E_\mathrm{XXZ}$ & -0.2707 & -0.3024 & -0.3160 & -0.3311 \\
\hline \hline $E^0(L)$, nn corr. & -0.2706 & -0.3018 & -0.3149 & -0.3292\\
\hline $\eta$ & 0.0837 & 0.1329 & 0.1481 & 0.1629 \\
\hline \hline $E^0(L)$, 3rd nn corr. & -0.2707 & -0.3024 & -0.3160 & -0.3309 \\
\hline $\eta_{10}$ & 0.0840 & 0.1353 & 0.1518 & 0.1670\\
\hline $\eta_{21}$ & 0.0025 & 0.0092 & 0.0129 & 0.0161\\
\hline $\eta_{30}$ & 0.0007 & 0.0037 & 0.0056 & 0.0073 \\
\hline
\end{tabular}
\end{minipage}
\caption{Lowest XXZ-model energies and parameters for the state
$e^{-L}|0\rangle$ on the square lattice. Reference energies are
the lowest values from the different methods listed in ref.
\cite{XXZGroundState}, except for $\alpha=1$, where the number is
the same as used in Table \ref{tab:E0}. \label{tab:E0L}}
\end{table}

Table \ref{tab:E0L} lists the results for $\alpha=0.5$, 0.8, 0.9
and 1 (in units of $J$). The row labelled 'Reference energy'
contains low numerical estimates as described in the caption. The
next row, labelled '$E_\mathrm{XXZ}$' are calculated using formula
(\ref{eq:HeisE0}). The row labelled '$E^0(L)$, nn corr.' gives the
results of expanding (\ref{eq:nncorrexpans}) to 7th order with the form of
$L$ given by (\ref{eq:LDef}) and calculating the minimum energy
with $\eta$ as a variational parameter. The values of $\eta$ are
listed below; they agree with the values found by
(\ref{eq:etaexact}) to within a few percent. To assess the
convergence of the expansion, it can be noted that the biggest
contribution to the energy from the 7th order term is -0.0002, for
$\alpha=1$. Hence, the accuracy is probably sufficient for
comparing the different approaches.

For all values of $\alpha$, it is seen that $E^0(L)$ is a poorer
estimate of the ground state energy than $E_\mathrm{XXZ}$, even
though the calculation of $E_\mathrm{XXZ}$ involves no free
parameters. In order to improve the variational result, I have
tried the calculation with further terms in the definition of $L$.
Since the expression (\ref{eq:LDef}) changes sign under exchange
of indices $j$ and $k$, it is not obvious how to have terms with
two spins from the same sublattice, as there is no natural way to
choose the sign for a given pair. Therefore, I have only added
terms between third-nearest neighbours. There are two types of 3rd
nearest neighbours: one type connected by a knight's move (two
bonds in a row followed by one perpendicular bond), with parameter
$\eta_{21}$, the other connected by three bonds in row with
parameter $\eta_{30}$. Following this notation, the amplitude of
terms involving nearest neighbours is called $\eta_{10}$. For the
calculation, all terms in (\ref{eq:nncorrexpans}) were expanded to 5th
order, and 6th and 7th order terms in $\eta_{10}$ were added. The
minimum energies obtained in this fashion are listed in the row
labelled '$E^0(L)$, 3rd nn corr.' and the parameters are provided
below. The lowering of the energy relative to the case with only
nearest-neighbour terms is modest, just bringing the result in
line with $E_\mathrm{XXZ}$ (except for $\alpha=1$). Improvements
from correlations between more distant spin pairs are probably
quite negligible, but lower energies might be obtained by adding 4-spin terms to $L$.

Although the energies of states of the type $e^{-L}|0\rangle$ are
further from the true ground state energies than the simple
estimate $E_\mathrm{XXZ}$ (not to mention the energies found by
more elaborate methods), the existence of low-lying states of this
new, simple form could still be of some interest. In particular, the
concept of delocalized spin operators might be a useful tool to account for quantum effects on magnetic order in more complex contexts, where the most accurate methods are difficult to apply. The transformation (\ref{eq:LTranspm}) can also be applied to localized fermions, with $s^+$ in (\ref{eq:LDef}) replaced by the creation operator $c^\dagger$ etc..

\section{Time evolution of spin operators in the Heisenberg model}
I now proceed to show how correlations between nearest neighbors as well as next-nearest neighbors can be applied in a calculation of the spin-wave dispersion of the $S=\frac{1}{2}$ Heisenberg model from the time evolution of the spin operators. The nearest-neighbor correlation is
\begin{eqnarray}
\langle \Sj \cdot \Sk \rangle & = & \langle 0 | e^{L} \Sj \cdot \Sk e^{-L}|0 \nonumber \rangle \\
& = & \langle 0 | e^{L} ( s^x_j x^x_k + s^y_j s^y_k + s^z_j s^z_k ) e^{-L}|0 \rangle
\label{eq:nncorrexpans}
\end{eqnarray}
This expression can be evaluated by series expansion as described in the previous section. In this section, the expansion is carried analytically to second order. The zero order term in the expansion is obviously $\langle 0 | s^z_j s^z_k | 0 \rangle = -1/4$. If we as before take $j$ to be on the 'up'-sublattice, the only contribution to first order comes from the commutator
\begin{equation}
2i\eta [s^x_j s^y_k - s^y_j s^x_k, s^x_j s^x_k + s^y_j s^y_k] = \eta s^z_k - \eta s^z_j = -\eta
\end{equation}
The only contribution to second order comes from commutating $s^z_j s^z_k$ with terms from $L$ that couple spin $j$ to a nearest neighbor $l$ different from $k$, or that couple spin $k$ to a nearest neighbor (of spin $k$)  different from $j$, {\it e.g.}
\begin{eqnarray}
\langle -\frac{(2\eta)^2}{2} [s^x_j s^y_l - s^y_j s^x_l,[s^x_j s^y_l - s^y_j s^x_l, s^z_j s^z_k]] \rangle \nonumber \\
= \eta^2 \langle -s^z_j s^z_k + s^z_k s^z_l \rangle
\label{eq:nn2ordercomm}
\end{eqnarray}
The expectation value in the ground state is $\eta^2/2$. Since there are $2(Z-1)$ such terms, the nearest neighbor correlation is to second order
\begin{equation}
\langle \Sj \cdot \Sk \rangle \approx -\frac{1}{4}-\eta+(Z-1)\eta^2
\label{eq:nn2ordereta}
\end{equation}
The ground state energy per bond is the nearest neighbor correlation multiplied by J, so the value of $\eta$ should be the one to give (\ref{eq:nn2ordereta}) its' minimal value. The minimal value is found for
\begin{equation}
\eta = \frac{1}{2(Z-1)}
\end{equation}
and the minimal value is
\begin{equation}
\langle \Sj \cdot \Sk \rangle = -\frac{1}{4}(1+\frac{1}{Z-1})
\label{eq:nn2orderZ}
 \end{equation}
which is just the first-order expansion of (\ref{eq:HeisE0}) except for the factor of $J$. So far nothing new. We can similarly calculate the correlation between next-nearest neighbors $j$ and $m$. The zero order term is 1/4, and the first order term is zero. The second order term has contributions similar to (\ref{eq:nn2ordercomm}), from the commutation of $s^z_j s^z_m$ with terms from $L$ that couple spin $j$ to a nearest neighbor $k$, or that couple spin $m$ to a nearest neighbor $l$, {\it e.g.}
\begin{eqnarray}
\langle -\frac{(2\eta)^2}{2} [s^x_j s^y_k - s^y_j s^x_k,[s^x_j s^y_k - s^y_j s^x_k, s^z_j s^z_m]] \rangle \nonumber \\
= \eta^2 \langle -s^z_j s^z_m + s^z_k s^z_m \rangle = -\eta^2/2
\label{eq:nnn2orderzzcomm}
\end{eqnarray}
 There are $2Z$ such terms. However, since $j$ and $m$ have a nearest neighbor $k$ in common, there is also a contribution from $s^x_j s^x_m+s^y_j s^y_m$ which first is commutated with a term in $L$ that couples $j$ to $k$ and then commutated with a term that couples $k$ to $m$, or vice versa, {\it e.g.}
\begin{eqnarray}
\langle -\frac{(2\eta)^2}{2} [s^x_m s^y_k - s^y_m s^x_k,[s^x_j s^y_k - s^y_j s^x_k, s^x_j s^x_m]] \rangle \nonumber \\
= -\frac{\eta^2}{2} \langle s^z_k s^z_m - s^z_j s^z_m\rangle = \frac{\eta^2}{4}
\label{eq:nnn2orderxxcomm}
\end{eqnarray}
There are 4 such terms, if we for simplicity ignore that some next-nearest neighbors have more than one nearest neighbor in common. We therefore arrive at the next nearest neighbor correlation
\begin{equation}
\langle \Sj \cdot \Sm \rangle = \frac{1}{4} - (Z-1)\eta^2 = \frac{1}{4}(1-\frac{1}{Z-1})
\label{eq:nnn2orderZ}
\end{equation}

The time evolution of ${\bf S}_j$ is governed by
\begin{equation}
\frac{\partial \Sj}{\partial t} = \frac{i}{\hbar} [H,\Sj] = -\frac{J}{\hbar} \sum_{k \in {\cal N}_j} \Sj \times \Sk
\label{eq:dSdt}
\end{equation}
In order to evaluate the time evolution of $\sum_k \Sj \times \Sk$ we need the following relation, which only holds for $S=\frac{1}{2}$:
\begin{equation}
[\Sj \cdot \Sk,\Sj \times \Sk] = \frac{i}{2}\Sk - \frac{i}{2} \Sj
\label{Eq:S1/2comm}
\end{equation}
By using this in addition to the more general commutation relation $[\Sl \cdot \Sj, \Sj \times \Sk] = i (\Sj \cdot \Sk) \Sl - i (\Sl \cdot \Sk) \Sj$ (for $l \neq k$) one obtains
\begin{eqnarray}
\nonumber \frac{\partial \sum_k \Sj \times \Sk}{\partial t} & = & \frac{i}{\hbar} [H, \sum_k \Sj \times \Sk] \\
\label{eq:dSixSjdt} & = & \frac{J}{2\hbar}  \sum_{k \in {\cal N}_j} \left\{ \left[ \Sj - \Sk \right] + \right. \\
\nonumber & & \frac{J}{\hbar} \sum_{l \in {\cal N}_j \atop l \neq k} \left[ \Sj (\Sk \cdot \Sl)-\Sl (\Sj \cdot \Sk) \right] +\\
\nonumber & & \frac{J}{\hbar} \sum_{m \in {\cal N}_k \atop m \neq j} \left .\left[ \Sm (\Sj \cdot \Sk) - \Sk (\Sj \cdot \Sm) \right] \right\}
\end{eqnarray}
Again, $k$ is a nearest neighbor to $j$; $l$ is a nearest neighbor to $j$, different from $k$; and $m$ is a nearest neighbor to $k$. If we replace the correlations with their expectation values in the antiferromagnetic ground state (\ref{eq:nn2orderZ}) and (\ref{eq:nnn2orderZ}) we find
\begin{equation}
\frac{\partial \sum_k \Sj \times \Sk}{\partial t}  \approx   \frac{J}{4\hbar}\frac{Z}{Z-1}  \sum_{k \in {\cal N}_j \atop m \in {\cal N}_k} \left[ \Sj - \Sm \right]
\label{eq:dSixSjdt_App}
\end{equation}
where the sum does not exclude $m=j$. Note that the actual eigenvalues of $\Sj \cdot \Sk$ are $-3/4$ and $1/4$. The value of  (\ref{eq:nnn2orderZ}) close to $1/4$ implies that the ground state is not too far from being an eigenstate of the next-nearest neighbor correlation, so the use of the expectation value is probably a safe approximation. The value of (\ref{eq:nn2orderZ}), on the other hand, is almost right between the two eigenvalues, so the ground state is decidedly not an eigenstate of the nearest-neighbor correlation. The use of the expectation value is a mean-field approach, and can be expected to carry some limitations.

\section{Raising and lowering operators for excitations}

I now define the following spin-wave operators
\begin{eqnarray}
\Wq & =& \frac{1}{\sqrt{N}} \sum_j e^{-i \ruj \cdot {\bf q}} \left(\Sj - i\beta({\bf q}) \sum_{k \in {\cal N}_j} \Sj \times \Sk \right)
\label{eq:Wqdef} \\
\Wdagq & =& \frac{1}{\sqrt{N}} \sum_j e^{i \ruj \cdot {\bf q}} \left(\Sj + i\beta({\bf q}) \sum_{k \in {\cal N}_j} \Sj \times \Sk \right)
\label{eq:Wdagqdef}
\end{eqnarray}
(where the  sum over $j$ runs over all spins) and solve the equations
\begin{eqnarray}
\frac{\partial \Wq}{\partial t} & = & - i \omega ({\bf q}) \Wq.
\label{eq:dWdt} \\
\frac{\partial \Wdagq}{\partial t} & = &  i \omega ({\bf q}) \Wq,
\nonumber
\end{eqnarray}
which imply that $\Wdagq$ increases the energy by $\hbar\omegaq$ and $\Wq$ decreases it by the same amount.  By virtue of  (\ref{eq:dSdt}), which is an exact relation, one obtains the amplitude of the cross product term
\begin{equation}
\beta({\bf q}) =   \frac{J}{\hbar \omega ({\bf q})}.
\label{eq:SWCrossAmplitude}
\end{equation}
Insertion of this relation into (\ref{eq:Wqdef}) and (\ref{eq:Wdagqdef}) reveals an analogy to the raising and lowering operators of the harmonic oscillator. Further applying (\ref{eq:dSixSjdt_App}), which is an approximation for the antiferromagnetic ground state, gives the approximate energies
\begin{equation}
\hbar \omega_\mathrm{AF} ({\bf q}) = \frac{Z}{2} \sqrt{Z/(Z-1)}\sqrt{J(0)^2-J({\bf q})^2}.
\label{eq:SWenergy}
\end{equation}
$J({\bf q})$ is defined by
\begin{equation}
J({\bf q}) = \frac{J}{Z} \sum_{k \in {\cal N}_j} e^{i (\ruj -\ruk) \cdot {\bf q}}
\end{equation}
where $\ruj$ is the position of spin $j$.
The spin-wave dispersion obtained is the well-known result, multiplied by a factor $\sqrt{Z/(Z-1)}$. For the square lattice this factor evaluates to $\sim 1.155$, which can be compared to the results of Zheng et al., who by series expansion calculate a factor between 1.09 and 1.19, depending on the position in the Brillouin zone \cite{Zheng2005}. 
The calculation above is explicitly for $S=\frac{1}{2}$ through the use of equation (\ref{Eq:S1/2comm}), so the Haldane conjecture is not violated.
 

It is easy to do the calculation for the ferromagnetic case, by inserting the ferromagnetic value +1/4 for both nearest-neighbor and next-nearest neighbor correlations into (\ref{eq:dSixSjdt}) and using the result instead of (\ref{eq:dSixSjdt_App}). One duly obtains the well-known ferromagnetic spin wave dispersion
\begin{equation}
\hbar \omega_\mathrm{FM} ({\bf q}) = \frac{Z}{2} \left( J({\bf q})-J(0) \right)
\label{eq:FMSWenergy}
\end{equation}
(note that $J$ is negative for the ferromagnet). It might be interesting to explore a mean field approach in which the spin waves influence each other through the effect on the nearest-neighbor and next-nearest neighbor correlations. On the other hand, it is at least for the ferromagnetic spin waves known that the interaction between spin waves with wave vectors ${\bf q}_1$ and ${\bf q}_2$ contains terms proportional to 
${\bf q}_1 \cdot {\bf q}_2$\cite{Dyson1956a}, which can not be approximated well by a mean field approach.

The definitions (\ref{eq:Wqdef}) and (\ref{eq:Wdagqdef}) apply to both the antiferromagnetic and ferromagnetic cases. The change of sign in the phase is allowed as long as $\omega ({\bf q}) =  \omega (-{\bf q})$ and ensures that combinations such as $\Wdagq \cdot\Wq$ do not cause displacement in reciprocal space.  Clearly, products of this type are constants of motion within the approximation developed here.


We can relate $\Wq$ and $\Wdagq$ to $H$ by noting that
\begin{equation}
\frac{1}{\sqrt{N}} \sum_j e^{ i \ruj \cdot {\bf q}} \Sj = \frac{1}{2} (\Wdagq + \Wnegq)
\end{equation}
and
\begin{equation}
\frac{i}{\sqrt{N}} \sum_j e^{ i \ruj \cdot {\bf q}} \sum_{k \in {\cal N}_j} \Sj \times \Sk = \frac{\hbar \omegaq}{2} (\Wdagq - \Wnegq).
\end{equation}
If we combine with the  relations
\begin{equation}
\Sj \cdot (\Sj \times \Sk) =  i\Sj \cdot \Sk
\end{equation}
and
\begin{equation}
(\Sj \times \Sk) \cdot \Sj = - i\Sj \cdot \Sk
\end{equation}
we obtain
\begin{eqnarray}
\sum_{\bf q} \hbar \omegaq  (\Wdagnegq - \Wq)  \cdot (\Wdagq + \Wnegq) & =  \nonumber \\
\frac{iJ}{N} \sum_{\bf q} \sum_j \sum_l \sum_{k \in {\cal N}_l} e^{ i (\ruj - \rul) \cdot {\bf q}}  (\Sl \times \Sk) \cdot \Sj & = \nonumber \\
J\sum_j \sum_{k \in {\cal N}_j} \Sj \cdot \Sk  
\end{eqnarray}
and
\begin{eqnarray}
\sum_{\bf q} \hbar \omegaq  (\Wdagnegq + \Wq)  \cdot (\Wdagq -\Wnegq) & =   \nonumber \\
\frac{iJ}{N} \sum_{\bf q} \sum_j \sum_l \sum_{k \in {\cal N}_l} e^{ i (\rul - \ruj) \cdot {\bf q}} \Sj \cdot (\Sl \times \Sk) & =  \nonumber \\
- J\sum_j \sum_{k \in {\cal N}_j} \Sj \cdot \Sk  &
\end{eqnarray}
This gives two different ways to express $H$. The most appealing expression is obtained by calculating the difference between the two, divided by two:
\begin{equation}
H = \sum_{\bf q} \hbar \omega ({\bf q}) \left( \Wdagq \cdot \Wq - \Wq \cdot\Wdagq \right)
\label{eq:HWcommu}
\end{equation}
Further investigation of the algebra of the $\Wdagq$ and $\Wq$ operators is required to evaluate the utility of this approach.

I am grateful to Kim Lefmann for illuminating discussions
and comments to the manuscript.

\end{document}